# Fabrication of Fiber-Reinforced Polymer Ceramic Composites by Wet Electrospinning

Yunzhi Xu[1], Junior Ndayikengurukiye[2], Ange-Therese Akono[1,2,*], Ping Guo[2,*]

[1] Department of Civil and Environmental Engineering, Northwestern University, Evanston, IL, USA
[2] Department of Mechanical Engineering, Northwestern University, Evanston, IL, USA
*Corresponding authors: Ping Guo ping.guo@northwestern.edu
Ange-Therese Akono: ange-therese.akono@northwestern.edu

**Abstract:**

We propose a novel approach of wet electrospinning to yield fiber-reinforced polymer ceramic composites, where a reactive ceramic precursor gel is used as a collector. We illustrate our approach by generating polyethylene oxide (PEO) fibers in a potassium silicate gel; the gel is later activated using metakaolin to yield a ceramic-0.5 wt% PEO fiber composite. An increase of 29% and 22% is recorded for the fabricated polymer ceramic composites in terms of indentation modulus and indentation hardness respectively. Our initial findings demonstrate the process viability and might lead to a potentially scalable manufacturing approach for fiber-reinforced polymer ceramic composites.

**Introduction**

Electrospinning is a versatile and efficient approach for generating micro-and nanofibers with an extremely high aspect ratio and surface area. The intrinsic wavy and spiral characteristics of electrospun fibers provide desirable attributes for toughening ceramic matrices. In traditional electrospinning, the fibers are usually collected as a non-woven mat using solid collectors. Instead, the adoption of a liquid bath collector has been exploited to either expedite the coagulation of fibers [1] or obtain fibers with specific structures [2]. Using electrospun fibers as nanoadditives has been explored primarily in polymer-based composites through film-stacking [3] or solution impregnation [4]. Short electrospun fiber reinforcement has been attempted through mechanical cutting [5] or ultrasonication [6] of electrospun nonwovens. However, these methods have not given full scope to the advantages of electrospun fibers. The compact nature of the non-woven mat obtained from traditional electrospinning substantially diminishes the flexibility of electrospun fibers, and the fibers cannot be dispersed inside the matrix. Meanwhile, the short electrospun fiber generation raises strict requirements in experiment devices and protocols and sometimes yields limited dispersibility [6].

The aforementioned obstacles in dispersing continuous electrospun nanofibers in composites can be alleviated using the wet electrospinning technique. Wet electrospinning, which adopts a liquid bath collector instead of a solid metal collector, provides a reliable approach for fabricating 3D porous fiber structures. Taskin et al. [7] has produced 3D microfibrous scaffolds using a grounded ethanol collector and demonstrated the advantages of the loosely packed fibrous structures in mimicking extracellular matrix compared to conventional 2D electrospun nonwovens. Sonseca et al. [8] has investigated the surface morphology of 3D scaffolds fabricated using wet electrospinning and 2D scaffolds using conventional electrospinning. The 3D scaffolds exhibited a 12% increase in open porosity, with each individual fiber possessing nanoporosity which facilitates cell infiltration. Chen et al. [9] has also demonstrated an increase of surface area to 6.45 $m^2/g$ of the 3D scaffold created by wet electrospinning. Despite its wide application in tissue engineering [10], wet electrospinning has not been explored in composite manufacturing, especially ceramic composites. The combination of electrospinning polymer nanofibers and a ceramic matrix can lead to great potential and versatility in creating organic/inorganic composites mimicking cortical bones [11].

In this letter, we propose a novel wet electrospinning approach, where a reactive ceramic precursor gel is used as the collector. The gel is later activated using metakaolin to yield fiber-reinforced polymer ceramic composites. The process will facilitate the random and uniform dispersion of electrospun nanofibers directly inside the matrix. The adoption of a liquid gel as the collector during electrospinning enables the direct infusion of nanofibers in an inorganic ceramic matrix at the same time when they are generated, which significantly reduces the time and effort required in conventional composite manufacturing methods.

**Methods**

**Wet Electrospinning of PEO Fibers**

Poly(ethylene oxide) (PEO) was chosen as the material of electrospun nanofibers due to its low toxicity and easy handleability. The 5 wt% PEO solution used for electrospinning was prepared by dissolving PEO powder (with a molecular weight of 600,000 g/mol) in deionized water and stirring at 30 ℃ using a magnetic stirrer for 12 h until full dissolution. The reactive ceramic precursor gel was a potassium silicate solution that was chosen for its versatility. The ceramic precursor was created by mixing fumed silica with potassium hydroxide pellets dissolved in deionized water using a magnetic stirrer. The mixture was allowed to sit on an orbital shaker for 24 h to ensure a homogeneous solution.

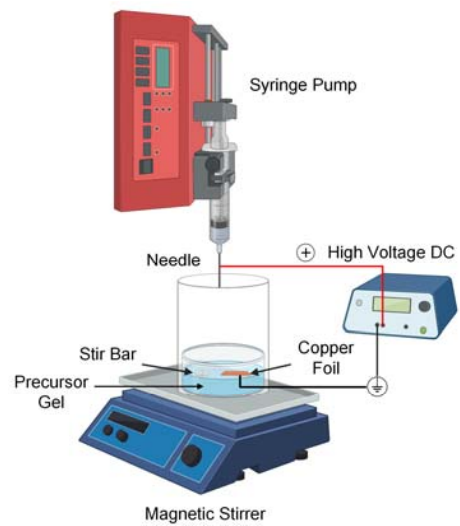

**Figure 1**: The experimental setup for wet electrospinning into a reactive ceramic precursor gel.

The schematic of the experimental setup is shown in **Fig. 1**. PEO solution was loaded in a 6 ml syringe with a stainless-steel needle of gauge 21 (inner diameter of 584 µm) and fed using a syringe pump (NE-300, New Era Pump Systems, USA). Positive DC high voltage (ranging from 0 to 11 kV) was generated by a high voltage amplifier (Trek 10/10B-HS, Advanced Energy, USA) and applied to the needle. The ceramic precursor was contained by a petri dish and served as the liquid gel collector for electrospun fibers. A small copper foil connected to the ground was inserted into the liquid gel and placed on the bottom of the petri dish. The ceramic precursor was stirred at 600 rpm with a magnetic stirrer to facilitate the fiber dispersion inside the liquid gel during collection. In order to avoid any disturbance of the electric field by other possible grounds, a plastic sheet was wrapped around the petri dish to isolate the electrospinning environment.

**Electrospun PEO Fiber-Ceramic Composite Synthesis**

30 g of electrospun PEO fiber-reinforced polymer ceramic was synthesized containing 0.5 wt% of PEO fibers. The electrospinning process took 3 hr to infuse 0.15 g of PEO fibers (3 g of 5 wt% PEO solution) into 19.68 g of ceramic precursor gel. After obtaining the ceramic precursor gel with uniformly dispersed nanofibers, the mixture was combined with 10.17 g of metakaolin, an activator, using a planetary centrifugal mixer (ARE-310, THINKY, USA) at 1200 rpm for 10 minutes and degassed at 1400 rpm for 5 minutes to form the fresh polymer-ceramic composite, potassium geopolymer slurry reinforced with PEO fibers. The slurry was then cured at 50 ℃ for 24 h on an orbital shaker running at 160 rpm to allow the further escape of macroscopic air bubbles.

**Results and Discussion**

**Dispersion of Electrospun PEO Fibers in Liquid Gel**

In principle, with the assistance of a magnetic stirrer, the electrospun fibers will disperse uniformly inside the liquid gel. However, the results suggest that PEO fibers tend to form agglomerations inside the ceramic precursor gel, as shown in **Fig. 2(a)**. Therefore, a further dispersion strategy was employed to achieve a more uniform dispersion of the fibers. The clusters were first untangled using an overhead stirrer (RW 20 digital, IKA, Germany) for 30 minutes at 1000 rpm. The mixture was then stirred overnight at 800 rpm using a magnetic stirrer to further break down the agglomerations. The dispersion result is shown in **Fig.**

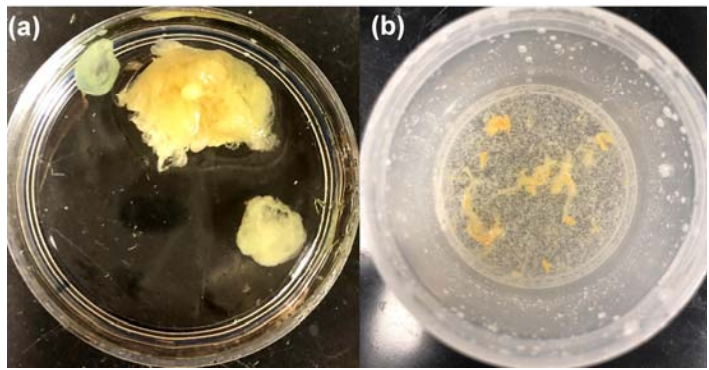

**Figure 2**: (a) As deposited and (b) dispersed electrospun PEO fibers inside the ceramic precursor gel.

**2(b)**, where the nanofibers were dispersed to a great extent except for some small clusters bound together by the unevaporated solvent.

**Microstructure of Hardened Electrospun Fiber-Reinforced Polymer-Ceramic Composites**

**Fig. 3(a)** displays the microstructure of the resulting ceramic-0.5 wt% eletrospun PEO fibers composite. A random distribution of the electrospun fibers is observed. The average fiber diameter is 5.66 ± 2.84 µm given by digital image analysis. The details are given in the Supplementary Material. **Fig. 3(b)** shows the existence of individual PEO fibers inside the matrix. The interface between the fiber and the matrix is blended, suggesting good bonding between PEO fibers and the matrix and the enhanced geopolymerization at the interface [12, 13]. This can be attributed to the surface nanoporosity exhibited in wet-electrospun fibers which facilitates the reaction and impregnation of the ceramic matrix [8]. **Fig. 3 (c, d)** displays a crack-bridging effect introduced by electrospun fibers, which has also been reported for geopolymer nanocomposites reinforced with carbon nanofibers [12] and serves as a major toughening mechanism for fiber-reinforced composites.

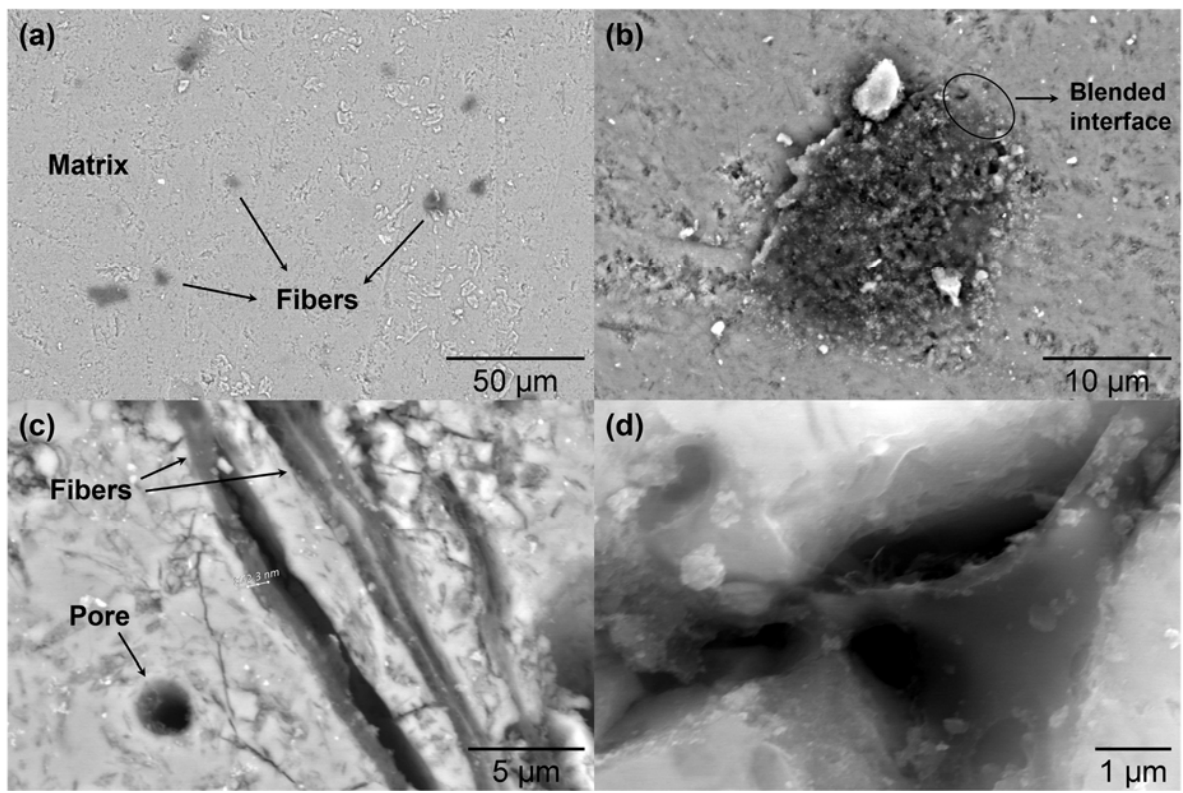

**Figure 3**: (a) Microstructure of hardened ceramic-0.5 wt% eletrospun PEO fibers; (b) individual PEO fiber inside the ceramic matrix; and (c, d) fracture micromechanisms of hardened nanocomposites.

**Mechanical Properties of Hardened Electrospun Fiber-Reinforced Polymer-Ceramic Composites**

**Fig. 4** displays the measured mechanical characteristics of the hardened ceramic-0.5 wt% electrospun PEO fiber composite as measured by microindentation, in comparison with the mechanical properties of a pure ceramic matrix. The detailed experimental procedure is given in the Supplementary Material. The maximum penetration depth is 7.03 µm, indicating that the indentation test is probing a local volume of a characteristic radius of 21 μm [15]. The distributions of both the indentation modulus and hardness exhibit a bell-shaped curve with a broad peak. The average value of the indentation modulus is $M = 10.72 \pm 0.68$ GPa, whereas the average value of the indentation hardness is $H = 524.13 \pm 49.67$ GPa. Thus, the variability in the indentation modulus is 6.3% and the variability in the indentation hardness is 9.5%. The high variability of the mechanical properties points to the heterogeneity of the microstructure introduced by electrospun fiber reinforcement. Therefore, in future studies, we will investigate ways to better control the microstructure using wet electrospinning.

Compared to a pure ceramic matrix fabricated according to the same protocol but without electrospun fibers, see **Fig. 4(a)-(b)**, the ceramic-0.5 wt% electrospun PEO fiber composite exhibits a 29% increase in indentation modulus and a 22% increase in indentation hardness. The enhancement in mechanical properties can be explained by the fact that electrospun fibers serve as both a catalyst and a reinforcing phase in the composite. During the curing of the composite, the PEO fiber is a catalyst for the geopolymerization reaction due to its high surface area [9, 14]. When the composite is hardened, the PEO fibers give rise to crack-bridging effects that strengthen the material. The increase in elasto-plastic characteristics shows that the process of using a precursor gel as a collector during the electrospinning process is efficient to yield performance-enhanced fiber-reinforced polymer-ceramic composites.

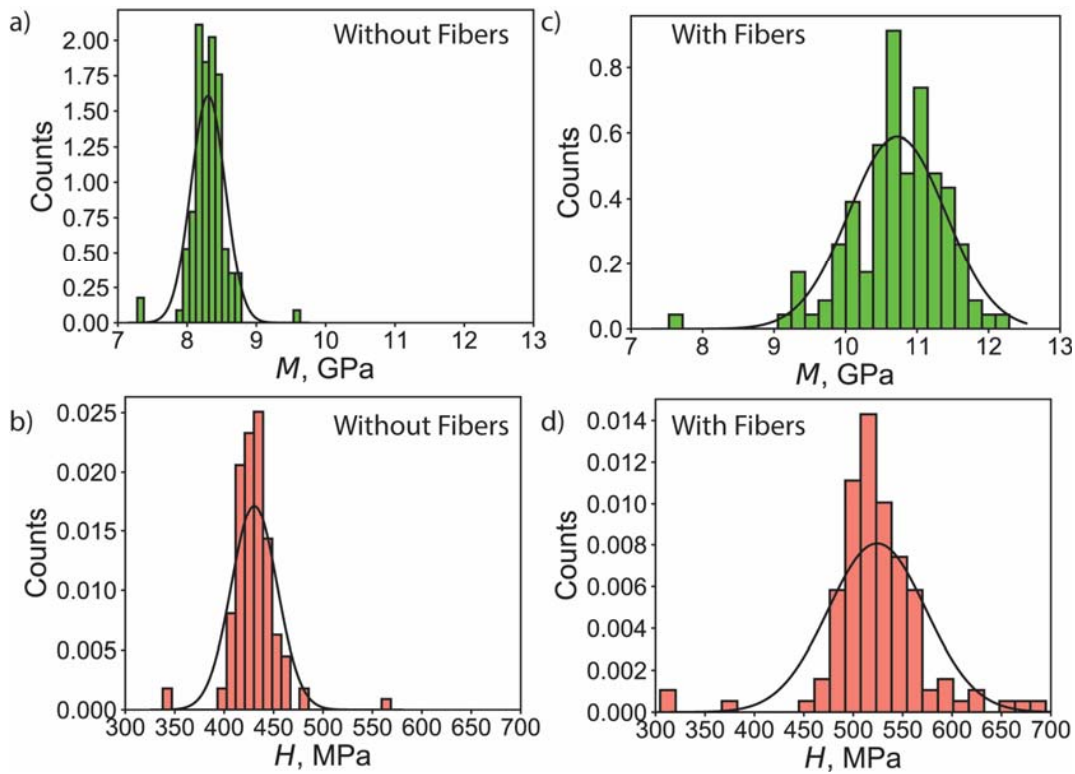

**Figure 4**: Mechanical properties of ceramic-0.5 wt% eletrospun PEO fiber composite compared to the pure unreinforced ceramic. (a)-(b) Pure unreinforced ceramic [12]; (c)-(d) ceramic-PEO fiber composite. $M$ is the indentation modulus, whereas $H$ is the indentation hardness.

**Conclusions**

We investigated a novel approach of wet electrospinning to yield performance-enhanced polymer-ceramic composites by electrospinning polymer fibers directly to a reactive ceramic precursor gel. Our initial results supported its feasibility by successfully fabricating fiber-reinforced ceramics by curing the ceramic precursor and fiber mixture with an activator agent after the wet electrospinning. An important concern was to homogeneously disperse the fibers and eliminate fiber clusters. The resulting hardened fiber-reinforced polymer-ceramic composite showed a random distribution of the PEO fibers with individual fiber well combined with the matrix. Mechanical characterization using microindentation tests revealed that 0.5 wt% electrospun PEO fibers were sufficient to yield a substantial increase of 29% and 22% in the indentation modulus and indentation hardness, respectively. This enhancement can be attributed to the fact that PEO fibers catalyze the ceramic precursor gel reaction due to their high surface area, as well as toughen the ceramic matrix through crack-bridging mechanisms. These initial findings demonstrate the process viability and might lead to a potentially scalable manufacturing approach for fiber-reinforced polymer ceramic composites.

**Acknowledgments and Funding**

This work is partially supported by AECC-BIAM (BIAM2020S1) till Jan 17, 2021.

# Supplementary Material

## Fabrication of Fiber-Reinforced Polymer Ceramic Composites by Wet Electrospinning

Yunzhi Xu[1], Junior Ndayikengurukiye[2], Ange-Therese Akono[1,2,*], Ping Guo[2,*]

[1] Department of Civil and Environmental Engineering, Northwestern University, Evanson, IL, USA

[2] Department of Mechanical Engineering, Northwestern University, Evanson, IL, USA

*Corresponding authors: Ping Guo ping.guo@northwestern.edu

Ange-Therese Akono: ange-therese.akono@northwestern.edu

## Experimental Procedure

### Environmental Scanning Electron Microscopy

The microstructure of the hardened electrospun PEO fibers/ceramic composite was characterized using optical microscopy and scanning electron microscopy. Optical microscopy analyses were conducted using a high-resolution Nikon transmitted-light optical microscope. Meanwhile, scanning electron microscopy was performed using an FEI Quanta 650 environmental scanning electron microscope in backscattered mode. The working distance was 10 mm. The accelerating voltage was 10-30 kV with magnification levels of ×1000-×2,5000.

### Digital Image Analysis

Digital Image Analysis was performed using the computer program ImageJ based on a grayscale SEM image of the hardened electrospun PEO fiber-reinforced ceramic composite, as shown in **Fig. S1**. Each fiber diameter is measured using the built-in line measurement function and the average fiber diameter is calculated based on the statistics obtained. The detailed fiber diameters measured are listed in Table S1.

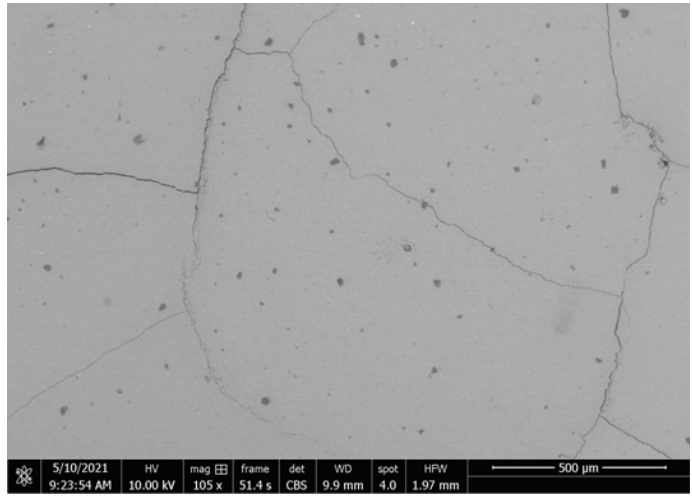

**Figure S1**: Microstructure of hardened ceramic-0.5 wt% eletrospun PEO fibers.

**Table S1**: Fiber Diameters measured using digital image analysis (µm).

| 11.985 | 4.646 | 4.556 | 9.158  | 9.225 | 9.112 | 2.038  | 8.376 |
|--------|-------|-------|--------|-------|-------|--------|-------|
| 3.866  | 4.556 | 7.759 | 4.126  | 3.757 | 5.763 | 3.222  | 4.075 |
| 3.222  | 5.195 | 3.919 | 2.882  | 2.657 | 6.346 | 3.285  | 9.665 |
| 3.866  | 5.94  | 7.204 | 5.032  | 3.222 | 7.759 | 3.47   | 9.579 |
| 1.441  | 3.919 | 3.222 | 4.691  | 3.645 | 3.285 | 10.329 | 6.94  |
| 5.313  | 6.346 | 5.505 | 5.195  | 7.088 | 3.285 | 11.616 | 5.195 |
| 6.94   | 3.47  | 2.038 | 12.692 | 2.657 | 5.195 | 12.511 | 2.657 |
| 8.645  | 4.556 | 7.865 | 4.126  | 3.222 | 3.285 | 5.505  | 1.822 |
| 4.51   | 2.657 | 5.155 | 7.204  | 3.222 | 5.763 | 14.48  |       |
| 2.038  | 10.39 | 6.079 | 5.94   | 6.849 | 6.113 | 7.346  |       |

**Micro-indentation Testing**

The elasto-plastic properties of the hardened electrospun fiber-reinforced polymer-ceramic composites were measured using micro-indentation testing. Micro-indentation consists of pushing a Berkovich diamond probe against the sample surface under a prescribed vertical force while measuring the penetration depth. The micro-indentation tests were conducted using an Anton Paar nanoHardness tester with a maximum vertical force of 500 mN, a holding phase of 30 s, and loading/unloading rate of 1,000 mN/min. A 11×11 micro-indentation grid was conducted with an inter-indent spacing of 100 µm, spanning an area of 1 mm$^2$. During each indentation test, the local indentation modulus $M$ and hardness $H$ were calculated by application of Oliver and Pharr's method [1, 2]:

$$H = \frac{P}{A}; M = \frac{\sqrt{\pi}}{2}\frac{S}{\sqrt{A}} \qquad (1)$$

Here, $S$ is the unloading indentation stiffness, whereas $A$ is the projected contact area, which is a function of the indenter geometry and of the penetration depth.

**References**

[1] Oliver, W. C., & Pharr, G. M. (1992). An improved technique for determining hardness and elastic modulus using load and displacement sensing indentation experiments. *Journal of materials research*, *7*(6), 1564-1583.

[2] Oliver, W. C., & Pharr, G. M. (2004). Measurement of hardness and elastic modulus by instrumented indentation: Advances in understanding and refinements to methodology. *Journal of materials research*, *19*(1), 3-20.